\documentclass[aps,pra,superscriptaddress,twocolumn]{revtex4}

\usepackage{graphicx} 
\usepackage{amsmath}
 
\newcommand{\beq}{\begin{equation}}
\newcommand{\enq}{\end{equation}}

\begin{document}

\title{Feshbach resonances  in an optical lattice}
\author{D.B.M. Dickerscheid}
\email{dickrsch@phys.uu.nl}
\affiliation{Institute for Theoretical Physics, Utrecht University, 
Leuvenlaan 4, 3584 CE Utrecht, The Netherlands}
\author{U. Al Khawaja}
\affiliation{Physics Department,United Arab Emirates University, 
P.O. Box 17551, Al-Ain, United Arab Emirates}
\author{D. van Oosten}
\affiliation{Institute for Theoretical Physics, Utrecht University, 
Leuvenlaan 4, 3584 CE Utrecht, The Netherlands}
\author{H.T.C. Stoof}
\affiliation{Institute for Theoretical Physics, Utrecht University, 
Leuvenlaan 4, 3584 CE Utrecht, The Netherlands}

\affiliation{Department of Physics and Astronomy, SUNY, Stony
Brook, NY 11794-3800, USA}

\date{\today}

\begin{abstract}
We present the theory for ultracold atomic gases in an optical
lattice near a Feshbach resonance. In the single-band
approximation the theory describes atoms and molecules which can 
both tunnel through the lattice.
Moreover, an avoided crossing between the two-atom and the
molecular states occurs at every site.
We determine the microscopic parameters of the generalized
Hubbard model that describes this physics, using the
experimentally known parameters of the Feshbach resonance in the
absence of the optical lattice. As an application we also
calculate the zero-temperature phase diagram of an atomic Bose
gas in an optical lattice.
\end{abstract}


\maketitle

\section{Introduction.} In the last few years there has been much
excitement in the field of ultracold atomic gases. To a large
extent this is due to two new experimental developments. The first
is the use of so-called Feshbach resonances in the collision of
two atoms, and the second is the use of an optical lattice. Both
developments have led to an unprecedented controle over the
physically relevant parameters of the atomic gas that can be used
to explore new strongly-correlated regions of its phase diagram.
In this paper we propose to combine these two developments and
study an atomic gas in an optical lattice near a Feshbach
resonance.

A more specific motivation for studying Feshbach resonances in an
optical lattice is that recently it has been shown that in an
atomic Bose gas near a Feshbach resonance a quantum phase
transition occurs between a phase with only a molecular
condensate (MC) and a phase with both an atomic and a molecular
condensate (AC+MC) \cite{leo,mathijs}. The experimental
observation of this quantum Ising transition is, however,
complicated by the fact that in a harmonic trap the fast
vibrational relaxation of Feshbach molecules consisting of two
bosonic atoms appears to prevent the creation of a molecular
condensate in that case \cite{MIT}. In an optical
lattice with a low filling fraction molecule-molecule and
atom-molecule collisions can essentially be neglected and we
expect this problem to be much less severe.

Having this particular application in mind, we from now on focus on
atomic Bose gases. However, our results can be immediately
generalized to the case of a two-component Fermi gas or even an
atomic Bose-Fermi mixture in an optical lattice. Moreover, we
consider only such low filling fractions that it is justified to
neglect the possibility of having three or more atoms per lattice
site. The reason for this restriction is that in this case we
have at most two atoms per site and the effect of the resonant
interactions between the atoms can be incorporated into the
theory exactly. The latter was shown previously to be very
important for arriving at a quantitatively accurate description of
a harmonically trapped atomic gas near a Feshbach resonance
\cite{review}. How this can be achieved also in an optical
lattice is discussed next.

\section{Generalized Hubbard model.} 
\begin{figure}
\includegraphics[angle=270,width=1.0\columnwidth]{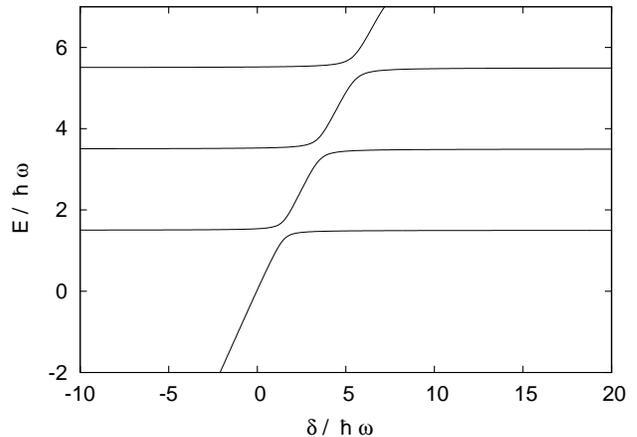}
\caption[Fig1]{
The relative energy levels of the atom-molecule system as a function of
 the detuning $\delta$. This figure was 
calculated for $g^{2} / \sqrt{2} \pi l^{3} (\hbar \omega)^{2} = 0.1$
}
\label{fig:selfe}
\end{figure}

We consider the
experimentally most interesting case of a deep optical lattice in
which the on-site potential is, for low energies, well approximated 
by an isotropic harmonic potential with energy splitting $\hbar\omega$ and the
tunneling energy $t_{\rm a}$ for atoms between sites obeys $t_{\rm a} \ll
\hbar\omega$.
For two atoms on a single site the two-channel Feshbach problem in 
the relative coordinate, after splitting off the center-of-mass motion,
 is then given by the Schr{\"o}dinger equation

\begin{equation}\label{eq:atmo}
\left(
\begin{matrix}
 H_{0} + V_{\rm aa}  
& V_{\rm am}  \\
 V_{\rm am}  &
\delta_B
\end{matrix}
\right)
\left(
\begin{matrix}
|\psi_{\rm a} \rangle \\
|\psi_{\rm m} \rangle
\end{matrix}
\right)
= E
\left(
\begin{matrix}
|\psi_{\rm a} \rangle \\
|\psi_{\rm m} \rangle
\end{matrix}
\right).
\end{equation}
Here the noninteracting atomic Hamiltonian is 
$H_{0} =  -\hbar^{2} \nabla_{\bf r}^{2}/m  +  m \omega^{2} {\bf r}^{2}/4 $.
The bare detuning is denoted by $\delta_{\rm B}$,
${\bf r}$ is the relative coordinate between the atoms and $m$ 
is the atomic mass. The nonresonant or background atom-atom interaction 
is $V_{\rm aa}$ and the 
atom-molecule coupling is denoted by $V_{\rm am}$.
In first instance only the relative part is relevant, 
since only this part is affected by 
the interactions between the atoms. The center-of-mass part determines the 
tunneling.
From Eq. (\ref{eq:atmo}) we obtain the following equation for the molecules 
\begin{eqnarray}\label{eq:2}
\langle \psi_{\rm m} |  V_{\rm am}  
\frac{1}{E -  H_{0} - V_{\rm aa} }
V_{\rm am}  |\psi_{\rm m} \rangle 
&=& E - \delta_{\rm B}, 
\end{eqnarray}
where $|\psi_{\rm m} \rangle$ is the bare molecular wavefunction.
Note that in the above we have implicitely taken the extend of this wavefunction
to be so small that its energy is not affected by the optical lattice,
which is well justified in practice.
Because for most atoms we also have that $|V_{\rm aa}| \ll \hbar \omega$, we can
neglect in the atomic propagator $V_{\rm aa}$ compared to $H_{0}$.
Moreover,
the eigenstates $|\phi_{n}\rangle$ of $H_{0}$ with energy
 $E_{n} = (2 n + 3/2) \hbar \omega$ that are relevant for an $s$-wave 
Feshbach resonance,
can be written in terms of 
the generalized Laguerre polynomials as
$\langle {\bf r}|\phi_{n} \rangle= e^{-{\bf r}^{2}/4 l^{2}} L_{n}^{1/2} ({\bf r}^{2}/2 l^{2})/ (2 \pi l^{2})^{3/4} [{L_{n}^{1/2}(0)} ]^{1/2}$. Here
 $l = \sqrt{\hbar/m \omega}$ is the harmonic oscillator length.
Using these states Eq. (\ref{eq:2}) can be rewritten as

\begin{eqnarray}
\sum_{n} 
\frac{\left|\langle \psi_{\rm m} | V_{\rm am}  | \phi_{n} \rangle \right|^{2}}
{E - E_{\rm n}} 
= E - \delta_{\rm B}.
\end{eqnarray}
Using also the usual pseudopotential approximation, we have that 
$\langle {\bf r} | V_{\rm am} | \psi_{\rm m} \rangle = \sqrt{2} g \delta({\bf r})$,
where the atom-molecule coupling 
$g = \hbar \sqrt{2 \pi a_{\rm bg} \Delta B \Delta \mu /m}$ 
depends on the background scattering length $a_{\rm bg}$, 
the width of the resonance $\Delta B$, and the difference in magnetic 
moments $\Delta \mu$ of the relevant Feshbach resonance \cite{review}.
From this we then find that the energy of the molecules obeys 
\begin{eqnarray}\label{eq:emo}
E - \delta_{B} &=& 2 g^{2} \sum_{m} \frac{\phi_{m}^{*}(0) \phi_{m}^{\phantom *} (0)}{E - E_{m}} 
\nonumber \\ &=& g^{2} \left[ 
\frac{G(E)}{\sqrt{2} \pi l^{3} \hbar \omega} - \lim_{r \rightarrow 0} 
\frac{m}{2 \pi \hbar^{2} r} \right].
\end{eqnarray}
The function $G(E)$ is the ratio of two gamma functions
$G(E) = \Gamma (- E/ 2 \hbar \omega + 3/4) / \Gamma (- E/2 \hbar \omega + 1/4)$.
The divergence in Eq.~(\ref{eq:emo}),
which was first obtained by Busch {\it et al.}\ in the context of a 
single-channel problem \cite{Wilkens}, can be dealt with by using 
the following  renormalisation procedure. 
The right-hand side of Eq.~(\ref{eq:emo}) can 
be interpreted as  the selfenergy of the molecules  
$\hbar \Sigma_{\rm m} (E)$.
The divergence in the selfenergy is  energy-independent 
and is related to an ultraviolet divergence that comes about because 
we have used pseudopotentials. 
To deal with this divergence we 
have to use the renormalized detuning instead of the bare detuning. The former is defined as 
$\delta = \delta_{B} - \lim_{{r} \downarrow 0} m g^{2} / 2 \pi \hbar^{2} {r}$,
where $\delta = \Delta \mu (B - B_{0})$ is determined by the experimental 
value of the magnetic field $B_{0}$ at resonance.
Note that, as expected, the required subtraction is exactly equal to the one needed
in the absence of the optical lattice.
In the latter case we have to subtract 
$2g^{2}  \int d {\bf k}~m / \hbar^{2} {\bf k}^{2} (2 \pi)^{3}$ \cite{review,Holland},
which can be interpreted as $\delta = \delta_{\rm B} - \lim_{{\bf r} \downarrow 0} 
2 g^{2} \int d {\bf k}~e^{i {\bf k}\cdot {\bf r} }m / \hbar^{2} {{\bf k}^{2}} (2 \pi)^{3}$.
In this manner we obtain the relative energy levels of the dressed molecules as
a function of the experimental detuning that is shown in 
Fig. \ref{fig:selfe}.

\begin{figure}
\includegraphics[angle=270,width=1.0\columnwidth]{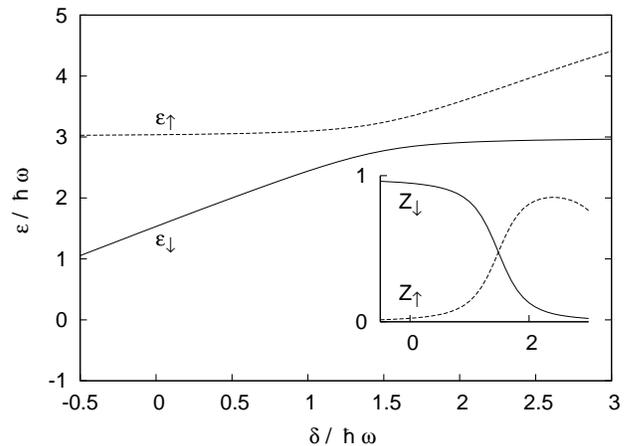}
\caption[Fig2]{
Details of the physical content of our theory.
We show the avoided crossing between the molecular level and the 
lowest two-atom trap state. The inset shows the probability $Z_{\sigma}$ as a function of 
the detuning $\delta$. This figure was 
calculated for  $g^{2} / \sqrt{2} \pi l^{3} (\hbar \omega)^{2} = 0.1$
Note that the center of mass contribution 
to the energy has been taken into account here.
}
\vspace{-.1cm}
\label{fig:avoided}
\end{figure}

From this figure we see that for very negative detuning the molecular state lies 
below the ground-state of the on-site microtrap and the bound-state energy is 
well approximated by the detuning. As it approaches the 
ground-state level of the trap there is an avoided crossing and as a result 
the lowest trap state is shifted upward. If the avoided crossings between 
the molecular level and subsequent trap states do not strongly overlap, 
the system can be well described by considering only the lowest trap state.
The overlap between the avoided crossings is determined by the strength of the
atom-molecule coupling and can be neglected if 
$g^{2}/\sqrt{2} \pi l^{3} (\hbar \omega)^{2} \ll 1$. 
In this paper we restrict ourselves
to a single-band approximation, although the generalization to the multi-band situation is 
straightforward.
This means that we only take into account the wavefunctions of the molecular state and 
the ground state of the on-site microtrap. 
In that case only two energy levels are of importance when there are 
two atoms on a lattice site. We denote these levels by 
$\epsilon_{\uparrow}$ and $\epsilon_{\downarrow}$ and their behaviour as
a function of detuning is shown in Fig. \ref{fig:avoided}.

The effective atom-molecule coupling in the 
optical lattice is given by $g' = g (\int d {\bf x} | \psi_{0} ( {\bf x}) |^{4})^{1/2}
= g/(2 \pi l^{2})^{3/4}$, where
$\psi_{0} ({\bf x})$ is the Wannier function in the lowest band of the optical lattice. 
The effective atom-atom interaction is now given  by
$U_{\rm eff} = U_{\rm bg} - 2 (g')^{2} / (\delta - 3 \hbar \omega/2)$, 
where the background on-site interaction strength 
$U_{\rm bg} = \left( 4 \pi a_{\rm bg} \hbar^{2} /m \right)
\int d {\bf x} |\psi_{0} ({\bf x}) |^{4}
= \sqrt{2/\pi} \hbar \omega \left( a_{\rm bg} / l \right)$. 
It is interesting to note that in order for the single-band 
approximation to be valid we do not need to have 
that $U_{\rm eff} \ll \hbar \omega$ because the on-site two-atom problem
has been solved exactly. 
In Fig. \ref{fig:avoided} we also show a close-up of the avoided crossing and the 
wavefunction renormalisation factors $Z_{\sigma}$ that give the amplitude of the 
closed channel part of the  molecules in the state $|\psi_{\sigma}\rangle$. 
Explicitely, we thus have in the single-band approximation that 
\begin{eqnarray}
&& | \psi_{\uparrow} \rangle 
= \sqrt{Z_{\uparrow}} |\psi_{\rm m} \rangle 
- \sqrt{1 - Z_{\uparrow}} |\psi_{0} \psi_{0} \rangle 
\nonumber \\
&& | \psi_{\downarrow} \rangle 
= \sqrt{Z_{\downarrow}} |\psi_{\rm m} \rangle 
+ \sqrt{1 - Z_{\downarrow}} |\psi_{0} \psi_{0} \rangle. 
\end{eqnarray}
The probability $Z_{\sigma}$ is determined 
by the selfenergy of the molecules  
through  the relation $Z_{\sigma} = 1/ (1 - \partial \hbar 
\Sigma_{\rm m} (E)/ \partial E )$ \cite{mathijs}. 
Note that in Fig. \ref{fig:avoided} the probability $Z_{\uparrow}$ already shows the effect of the avoided crossing
at a detuning of about $3\hbar \omega$. As long as the single-band approximation is valid this will, however, not 
affect any of the results because the two-atom state that is involved in this avoided crossing will not be populated.

Combining the above we thus find a generalized Hubbard Hamiltonian that is given by

\begin{widetext}
\begin{eqnarray}\label{eq:bh}
H &=&
-t_{a} \sum_{\langle i,j \rangle} a^{\dagger}_{i} a^{\phantom \dagger}_{j}
-t_{m} \sum_{\sigma} \sum_{\langle i,j \rangle} b^{\dagger}_{i,\sigma} b^{\phantom \dagger}_{j,\sigma}
\nonumber  + \sum_{\sigma} \sum_{i} 
\left( \epsilon_{\sigma} - 2 \mu \right) b^{\dagger}_{i,\sigma} b^{\phantom \dagger}_{i,\sigma}
 + \sum_{i} (\epsilon_{\rm a} -\mu ) a^{\dagger}_{i} a^{\phantom \dagger}_{i}
\nonumber   +
\frac{U_{\rm bg}}{2} \sum_{i}
a^{\dagger}_{i}
a^{\dagger}_{i}
a^{\phantom \dagger}_{i}
a^{\phantom \dagger}_{i}
\nonumber \\ && + g' \sum_{\sigma} \sum_{i} \sqrt{Z_{\sigma}} \left(
b^{\dagger}_{i,\sigma} a^{\phantom \dagger}_{i} a^{\phantom \dagger}_{i}
+ a^{\dagger}_{i} a^{\dagger}_{i} b^{\phantom \dagger}_{i,\sigma} \right).
\end{eqnarray}
\end{widetext}
Here $t_{a}$ and $t_{m}$ are the tunneling amplitudes for the atoms and the molecules, respectively,
and $\langle i,j \rangle$ denotes a sum over nearest neighbours. 
The operators $a^{\dagger}_{i}$, $a^{\phantom \dagger}_{i}$ correspond 
to the creation and annihilation operators of a single atom
at site $i$ respectively.
The operators $b^{\dagger}_{i,\sigma}$, $b^{\phantom \dagger}_{i,\sigma}$ 
correspond  to the creation and annihilation operators of the 
dressed molecules at site $i$ respectively.
Also $\epsilon_{\rm a} = 3 \hbar \omega/2$
is the on-site energy of a single atom.
In the tight-binding limit the hopping amplitudes can be conveniently expressed in terms of 
the lattice parameters as \cite{vanOosten2001a}
\begin{equation}
t_{\rm a,m} = \frac{\hbar \omega}{2} \left[ 1 - \left( \frac{2}{\pi} \right)^{2} \right]
\left( \frac{\lambda}{4 l_{\rm a,m}} \right)^{2} e^{-(\lambda/4 l_{\rm a,m})^{2} }.
\end{equation}
Here $\lambda$ is the wavelength of the light used to create the optical lattice and
 $l_{\rm m}\sqrt{2} = l_{\rm a} = l$.
Note that, as expected, we have that $t_{\rm m} \propto t_{\rm a}^{2} / \hbar \omega \ll t_{\rm a}$. 
Note also that our harmonic approximation to the on-site potential in 
principle slightly underestimates the hopping parameter.
A more accurate determination of these parameters would involve the
calculation of the appropriate Wannier functions.
The chemical potential $\mu$ is added because we perform 
our next calculations in the grand-canonical ensemble. 

\section{Phase diagram.} To find the mean-field phase diagram of a Bose gas in 
an optical lattice, we consider at sufficiently negative detuning the 
phase with only a Bose-Einstein condensate of molecules and perform a quadratic expansion  of 
the Hamiltonian in the 
fluctuations of the molecular annihilation operator $b_{{\bf k},\sigma}$ around the nonzero expectation value 
$\langle b_{{\bf k},\sigma} \rangle = \sqrt{n_{\rm mc}} \delta_{{\bf k},{\bf 0}} \delta_{\sigma,\downarrow}$. 
The effective Hamiltonian is then diagonalized by a Bogoliubov transformation and from the result we determine 
the equation of state of the gas as a function of the detuning $\delta$ and the 
temperature  $T \equiv 1/k_{B} \beta$. For the equation of state
for the total filling fraction
 we find (cf. Ref. \cite{mathijs}) $n = n_{\rm a} + 2 \sum_{\sigma} n^{\sigma}_{\rm m}$
with the molecular filling fractions obeying

\begin{eqnarray}
n^{\downarrow}_{\rm m} &=&  n_{\rm mc} + \frac{1}{N_{s}} \sum_{\bf k \neq 0} \frac{1}{e^{\beta \hbar 
\omega_{{\bf k},\downarrow}} - 1}, \nonumber \\ 
n^{\uparrow}_{\rm m} &=&   \frac{1}{N_{s}} \sum_{\bf k } \frac{1}{e^{\beta \hbar 
\omega_{{\bf k},\uparrow}} - 1},
\end{eqnarray}
and the atomic filling fraction
\begin{eqnarray}
n_{\rm a} &=& \frac{1}{N_{s}} \sum_{\bf k } \left\{ \frac{2 \epsilon^{\rm a}_{\bf k} - \epsilon_{\rm m} }{2 \hbar \omega_{\bf k}} 
\frac{1}{e^{\beta \hbar \omega_{\bf k}} -1}  \right. \nonumber \\ && \left.
+ \frac{2 \epsilon^{\rm a}_{\bf k} - \epsilon_{\rm m} - 2 \hbar 
\omega_{\bf k}}{4 \hbar \omega_{\bf k}} \right\}.
\end{eqnarray}
Moreover, we have that $N_{s}$ is the total number of sites in the lattice,
$\epsilon^{\rm a}_{\bf k} = -2 t_{\rm a} \sum_{j=1}^{3} \cos{\left( k_{j} \lambda/2 \right)} + \epsilon_{\rm a}$,
$\epsilon^{\rm m}_{{\bf k},\sigma} = -2 t_{\rm m} \sum_{j=1}^{3} \cos{\left( k_{j} \lambda/2 \right)} + \epsilon_{\sigma}$, and
 $\hbar \omega_{{\bf k},\sigma} =  \epsilon^{\rm m}_{{\bf k},\sigma}  + \epsilon_{\rm m}$ 
is the molecular dispersion.
Likewise we find that 
$\hbar \omega_{\bf k} = [\left( \epsilon^{\rm a}_{\bf k} -  \epsilon_{\rm m}/2 \right)^{2} 
- 4 g'^{2} Z_{\downarrow} n_{\rm mc}]^{1/2}$  is the atomic Bogoliubov dispersion with $\epsilon_{\rm m} = \epsilon_{\downarrow} - z t_{\rm m}$
equal to twice the chemical potential  and $z$ is the number of nearest neighbours.
 
\begin{figure}
\includegraphics[angle=270,width=1.0\columnwidth]{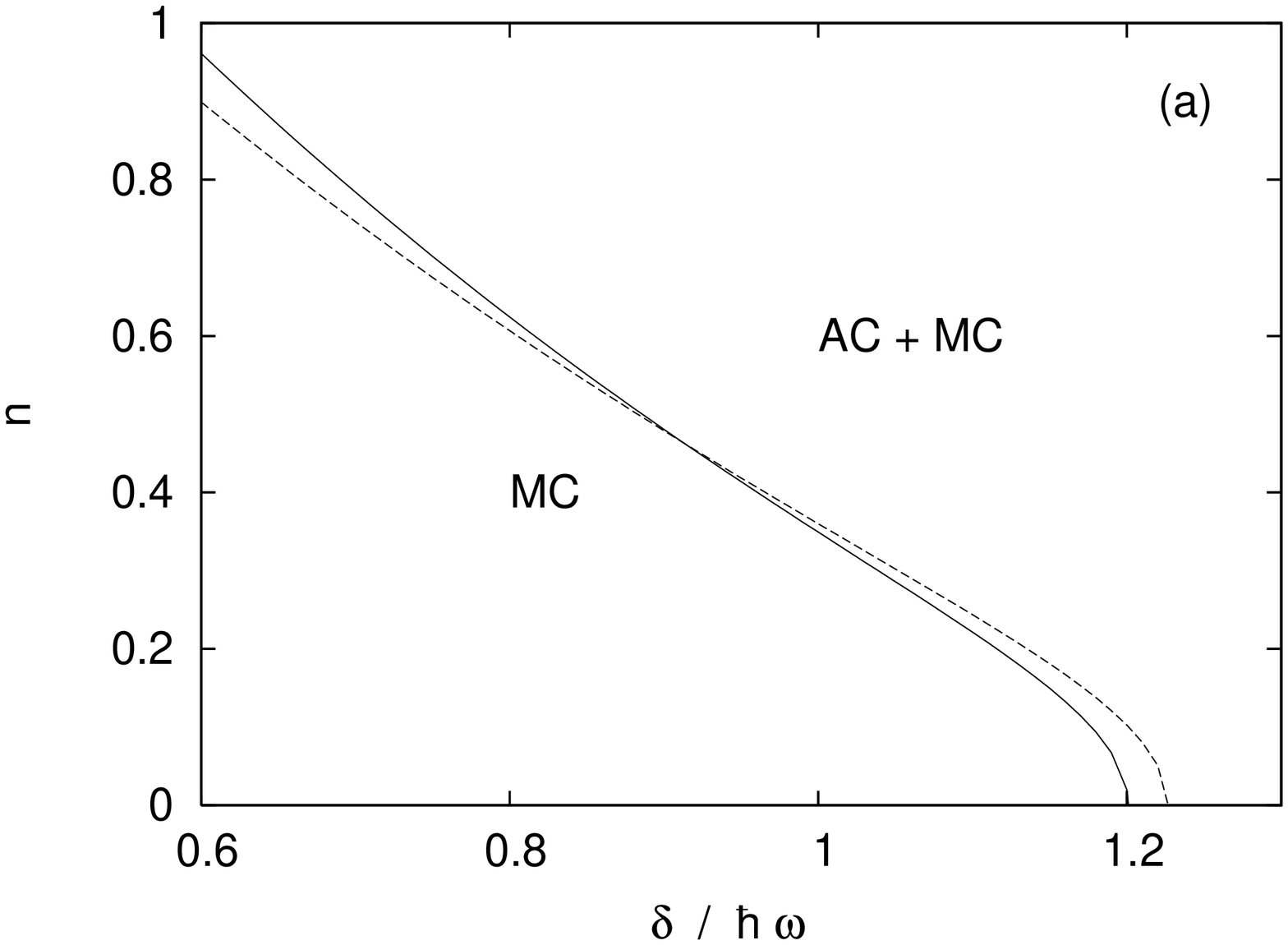}
\includegraphics[angle=270,width=1.0\columnwidth]{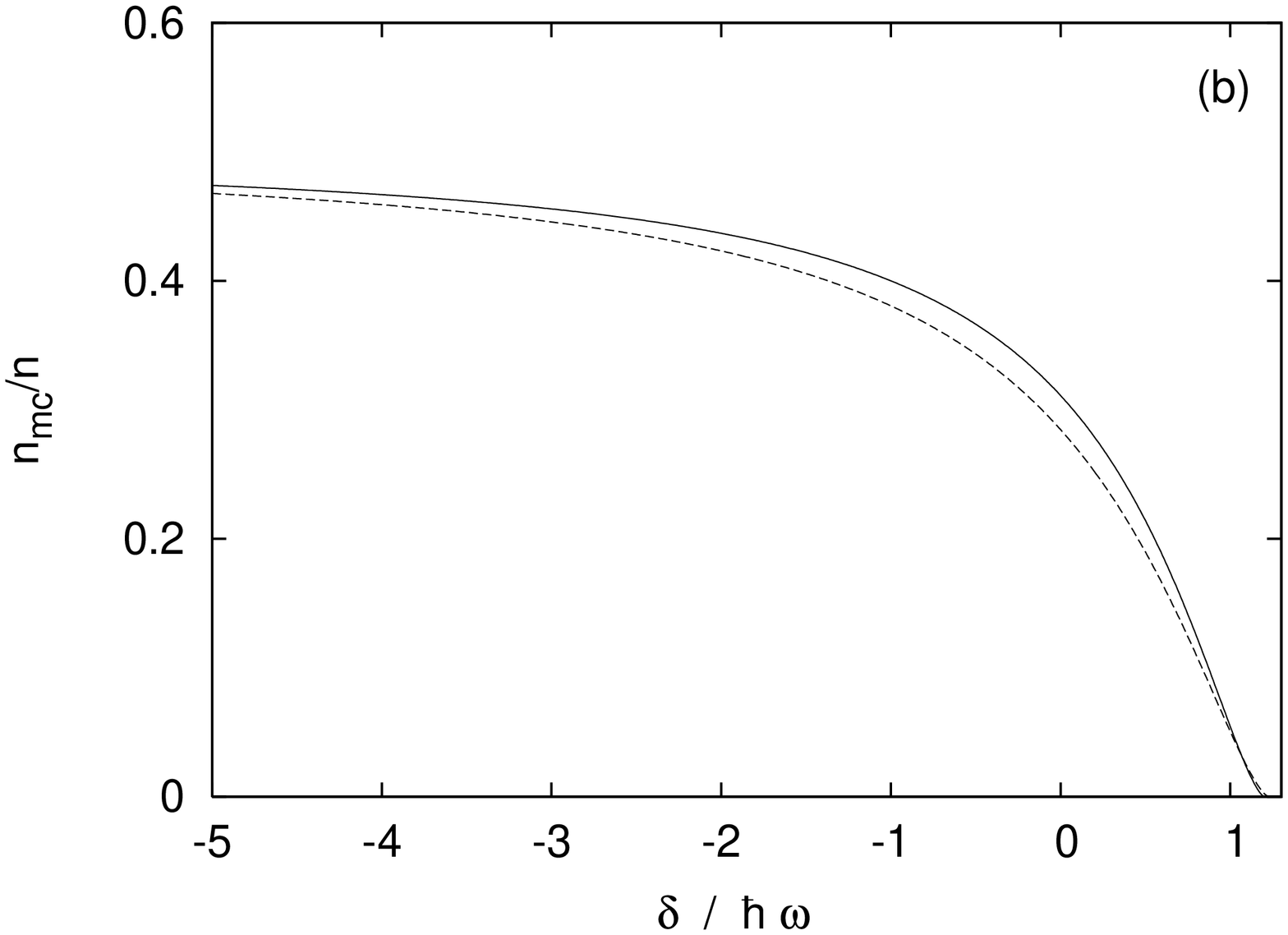}
\vspace{-.1cm}
\caption[Fig3]{
Zero temperature phase diagram as a function of the filling fraction per site and 
the detuning $\delta$ in units of $\hbar \omega$. The 
different curves that separate the MC and the AC+MC phases
correspond to values of $g'/\hbar \omega = 0.10$ (full curve) 
and $g'/\hbar \omega = 0.12$ (dashed curve) respectively. 
In both cases we have taken $\omega$ to be $10^{4}$~\mbox{rad/s}.
}
\label{fig:zeroTphasediag}
\end{figure}

The critical temperature for the Bose-Einstein condensation of the molecules follows from
the condition $n_{\rm mc} = 0$. The location of the Ising quantum phase transitions 
follows from the zero-momentum instability in the atomic Bogoliubov dispersion  when 
the detuning $\epsilon_{\rm m} = - 4 g' \sqrt{Z_{\downarrow}} \sqrt{n_{\rm mc}} + 2 \epsilon_{\rm a} - 2 z t_{a}$.
In Fig. \ref{fig:zeroTphasediag}a we show the results
for this condition as a function of the total filling fraction and detuning.
Note that in the limit of vanishing density the quantum critical point is determined by the 
ideal gas condition for Bose-Einstein condensation, i.e., $\mu = \epsilon_{\rm m} /2 = \epsilon_{\rm a} - z t_{\rm a}$.
From this condition it follows that for low enough filling fractions 
the location of the quantum phase transition
shifts to higher detuning with increasing strength of the atom-molecule coupling. 
On the other hand at large negative detuning
a larger value of the atom-molecule coupling implies a larger quantum depletion and hence a smaller molecular condensate fraction.
This effect shifts the Ising transition to
lower detuning.

For completeness we would like to point out that at $n=1$ the phase diagram can also contain 
a Mott-insulator phase \cite{yale}. This phase can occur at sufficiently large positive 
detuning such that $U_{\rm eff} / z t_{\rm a} \geq 3 + 2 \sqrt{2}$ 
\cite{vanOosten2001a,Dickerscheid2003a}. In contrast to the quantum Ising transition,
this transition has already been observed experimentally by Greiner {\it et al.} \cite{Greiner} 
after the theoretical prediction by Jaksch {\it et al.} \cite{Jaksch1998a}.
Its existence does not rely on the presence of the Feshbach resonance and we, therefore,
have not included it in the phase diagram in Fig. \ref{fig:zeroTphasediag}.
It is important to realize that  this Mott insulator can only exist for repulsive interactions between the atoms,
which requires $U_{\rm bg}$ to be positive.

\section{Conclusions and Discussion.} 

In summary, we have shown how to
 determine the microscopic parameters 
of the generalized Hubbard model in Eq.(\ref{eq:bh}) that describes the physics of resonantly-interacting 
atoms in an optical lattice, using the experimentally known parameters of 
the Feshbach resonance in the absence of the optical lattice. 
As an application we also calculated the zero-temperature phase diagram of an atomic Bose
gas in an optical lattice in the single-band approximation.
 By using an optical lattice one can suppress
three-body recombination processes that lead to a fast decay of the molecular condensate.

For the single-band approximation to be valid the atom-molecule
coupling constant $g$
 has to be small enough such that the avoided crossings between subsequent bands 
do not overlap with each other. In some cases, however, this coupling constant 
can be too large for realistic conditions and the single-band approximation will not 
hold anymore. In those cases we have to include higher-lying 
two-atom states. 
To avoid this complication the atom-molecule coupling  $g$ can be made smaller 
by using a more narrow Feshbach resonance or by using 
two-photon Raman transitions to convert atoms into molecules \cite{Bloch2004}.
Inclusion of higher-lying two-atom states is, however, easily achieved in our theory by adding
more atomic and molecular states into the generalized Hubbard model.
In principle, we have to add several dressed molecular states $|\psi_\sigma \rangle$
 for each additional
atomic band that is required for a sufficiently accurate description of the atomic gas
in the optical lattice. More precisely, for $M$ atomic bands we need to include $M^{2} + 1$ dressed molecular
states into the theory.

This work is supported by the Stichting voor Fundamenteel Onderzoek der 
Materie (FOM) and by the Nederlandse Organisatie voor 
Wetenschaplijk Onderzoek (NWO).
We are very grateful for helpful discussions with Tom Bergeman, Wolfgang Ketterle, and Martin Wilkens.

\bibliographystyle{apsrev}

\end{document}